\documentclass[aps,prl,twocolumn,showpacs,superscriptaddress,tightenlines,preprintnumbers,sort&compass,amsmath,amssymb,floatfix]{revtex4}

\usepackage{comment}
\usepackage{amssymb}
\usepackage{amsmath}
\usepackage[dvipsnames,usenames]{color}
\usepackage{graphicx,graphics}
\usepackage{amssymb}

\newcommand\ket[1]{\ensuremath{|#1\rangle}}
\newcommand\bra[1]{\ensuremath{\langle#1|}}

\begin{document}
\title{ Efficient tomography of quantum processes}
\author{Xiang-Bin Wang}
\affiliation{Department of Physics and  the Key Laboratory of Atomic
and Nanosciences, Ministry of Education, Tsinghua University,
Beijing 100084, China}
\affiliation{Advanced Science Institute,
RIKEN, Wako-shi, Saitama, 351-0198, Japan }
\author{Jia-Zhong Hu}
\affiliation{Department of Physics and  the Key Laboratory of Atomic
and Nanosciences, Ministry of Education, Tsinghua University,
Beijing 100084, China}
\author{Zong-Wen Yu}
\affiliation{Department of Physics and  the Key Laboratory of Atomic
and Nanosciences, Ministry of Education, Tsinghua University,
Beijing 100084, China}
\author {Franco Nori}
\affiliation{Advanced Science Institute, RIKEN, Wako-shi, Saitama,
351-0198, Japan } \affiliation{Physics Department,The University of
Michigan, Ann Arbor, Michigan 48109-1040, USA }

\begin{abstract}
We show with explicit formulas that one can completely identify an
unknown quantum process with only one weakly entangled state; and
identify a quantum optical Gaussian process  with either one
two-mode squeezed state or a few different coherent states. In
tomography of a multi-mode process, our method reduces the number of
different test states exponentially compared with existing methods.
\end{abstract}


\pacs{ 03.65.Wj, 42.50.-p, 42.50.Dv
} \maketitle

{\em Introduction.---}   One of the basic problems of quantum
physics is to predict the evolution of a quantum system under
certain conditions. For an isolated system with a known Hamiltonian,
the evolution is characterized by a unitary operator determined by
the Schr\"odinger equation. However, the system may interact with
its environment, and the total Hamiltonian of the system plus the
environment is in general not completely known.
The evolution can then be regarded as  a ``black-box
process"~\cite{qpt} which maps the input state into an output state.
An important problem here is how to characterize an unknown process
by testing the black-box with some specific input states, i.e.,
quantum process tomography (QPT)~\cite{qpt}.

Any physical process can be described by a completely positive map
$\varepsilon$. Such a process is fully characterized if the
evolution of any input state $\rho_{\rm in}$ is predictable:
$\rho_{\rm out}=\varepsilon (\rho_{\rm in})$. In general, QPT is
very difficult to implement in high dimensional spaces,  and, more
challengingly, in an infinite dimensional space, such as  a Fock
space. Recently, Ref.~\cite{calgary} showed QPT in Fock space, for
Continuous Variable (CV) states. Two conclusions~\cite{calgary} are:
($i$) If
 the output states of {\em all} coherent input states are known, then one
can predict the output state of any input state; ($ii$) By taking
the ``photon-number-cut-off approximation", one can then
characterize an unknown process with a finite number of different
input coherent states (CSs).

Here we study QPT using a different approach. Based on the idea of
 isomorphism~\cite{liusky}, and using the standard
$Q$-representation in quantum optics, we show, with explicit
formulas, that one can complete QPT with either only one weakly
entangled state for any quantum process, or only a few CSs for
quantum optical Gaussian processes. The method described here has
several advantages. First, it presents explicit formulas {\em
without} any approximation, such as the photon-number-cut-off
approximation. Second, it requires only one or a few different
states to characterize a process, rather than {\em all} CSs. Third,
for multi-mode Gaussian process tomography, the number of input CSs
increases {\em polynomially} with the number of modes, rather than
exponentially. Fourth, it uses $Q$-functions only, which is always
well-defined for {\em any} state without any higher order
singularities in the calculation.

{\em Isomorphism and process tomography with one weakly entangled
state.---} Define  $|\Phi^+\rangle=\sum_{k=0}^{s-1}|kk\rangle$ as
the $s$-level maximally entangled state in the composite space of
modes $a$ and $b$. (Here $|\Phi^+\rangle$ is not normalized, because
this simplifies the calculations below). Assume now that the process
$\varepsilon$ acts on mode $b$. Using isomorphism~\cite{liusky}, if
the state $\rho_{\varepsilon}=I\otimes \varepsilon
(|\Phi^+\rangle\langle \Phi^+|)$ is known, we shall know $
\varepsilon(\rho_{{\rm in}})$ for {\em any} single-mode input state
$\rho_{{\rm in}}$ on mode $b$.
 Consider a single-mode
input state on mode $b$,  $|\psi(\{c_k\})\rangle=\sum_k
c_k|k\rangle$. Obviously it can be written as
\begin{eqnarray}\label{00}
\left(|\psi\rangle\langle\psi|\right)_b ={_a}\langle \psi^*|
\Phi^+\rangle\langle \Phi^+|\psi^*\rangle_a\nonumber\\={\rm
tr_a}\left(|\psi^*\rangle\langle\psi^*|\otimes I\cdot
|\Phi^+\rangle\langle \Phi^+|\right) \end{eqnarray} and
$|\psi^*\rangle_a=\sum_k c_k^*|k\rangle_a$, is a single-mode state
on mode $a$ (sometimes we omit the subscript $a$ or $b$ for
simplicity). The output for any initial state $\rho_{{\rm
in}}=\left(|\psi\rangle\langle \psi|\right)_b$
 is
\begin{equation}\label{01}
\rho_{{\rm out}}=\varepsilon \left\{{\rm
tr_a}\left[\left(|\psi^*\rangle\langle \psi^*|\right)\!_a\otimes
I_b\cdot \left(|\Phi^+\rangle\langle
\Phi^+|\right)\!_{ab}\right]\right\}
\end{equation}
Since the partial trace and the map $\varepsilon$ are taken in
different subspaces, their orders can be exchanged. Thus
\begin{eqnarray}\label{e0}
\rho_{{\rm out}}={_a\langle} \psi^*| I\otimes
\varepsilon(|\Phi^+\rangle\langle
\Phi^+|)|\psi^*\rangle_a={_a\langle} \psi^*|
\rho_{\varepsilon}|\psi^*\rangle_a\nonumber\\={\rm tr_a}\left(
|\psi^*\rangle\langle\psi^*|\otimes I\cdot \rho_\varepsilon \right).
\end{eqnarray}
Equation (\ref{e0}) predicts the output state of any input state of
an unknown process, given   $\rho_{\varepsilon}$. However,
generating the maximum entangled state $|\Phi^+\rangle$ is
technologically difficult, especially when $s$ is large. Moreover,
for the case of CV states in Fock space, $s$ is infinite and the
maximum entanglement does not physically exist. Therefore,  we
cannot really test
 a process with $|\Phi^+\rangle$ in Fock space.
However, we can first test a process with some other
easy-to-manipulate states,  and then deduce the state
$\rho_\varepsilon$.
 For example,  one can test the one-sided map $I\otimes \varepsilon $
with an arbitrary non-maximally entangled state
$|\phi(\{r_k\})\rangle=\sum_{k=0}^{s-1} r_k|kk\rangle$, if $r_0\cdot
r_1\cdots r_{s-1}\not= 0$. Denoting the output state as
$\Omega_{\{r_k\}}$, we have
\begin{equation}\label{e1}
\Omega_{\{r_k\}}= I\otimes \varepsilon
(|\phi(\{r_k\})\langle\phi(\{r_k\})|).
\end{equation}
 On the other hand, we know that
$|\phi(r_k)\rangle = \hat T(\{r_k\})\otimes I |\Phi^+\rangle$;  and
$\hat T(\{r_k\})$ is a projection operator defined as $\hat T
(\{r_k\})=\sum_{k=0}^{s-1} r_k|k\rangle\langle k|  $. Since $\hat T
\otimes I$ and the one-sided map $I\otimes \varepsilon $ { commute},
then Eq.~(\ref{e1}) can be written as
\begin{equation}
\Omega_{\{ r_k \}}=\hat T(\{r_k\}) \otimes I\left\{ I\otimes
\varepsilon \left[|\Phi^+\rangle\langle \Phi^+\right]\right\}\hat
T(\{r_k\})\otimes I
\end{equation}
which gives rise to
\begin{equation}\label{e2}
\rho_{\varepsilon}=\hat T^{-1}(\{r_k\}) \otimes I\,
\Omega_{\{r_k\}}\, \hat T^{-1}(\{r_k\}) \otimes I.
\end{equation}
Here $\hat T^{-1}(\{r_k\})$ is defined as $\hat T(\{r_k^{-1}\})$. If
we test the one-sided map $I\otimes \varepsilon $ with the limited
entangled state $|\phi(\{r_k\})\rangle$ and we find that the outcome
state is $\Omega_{\{r_k\}}$, then, using Eq.~(\ref{e2}) we can
determine
 the output state when the input state is $\ket{\Phi^+}$.
We can then use Eq.~(\ref{e0}) to predict the output state of {\em
any} single-mode input state on mode $b$. Explicitly, if the input
state is $|\psi(\{c_k\})\rangle=\sum_k c_k|k\rangle$, the output
state becomes
\begin{equation}\label{ddim}
\rho_{\psi}={\rm tr_a}\left[ |\psi ({c_k^*}/{r_k})\rangle\langle
\psi ({c_k^*}/{r_k})|\otimes I \cdot \Omega_{\{r_k\}}\right]
\end{equation}
where $|\psi ({c_k^*}/{r_k})\rangle=\sum_k ({c_k^*}/{r_k})
|k\rangle$.

For a state in  Fock space, $s$ is infinite and $|k\rangle$ is a
Fock state which can be generated by the creation operator
$a^\dagger$ on the vacuum state $|0\rangle$.  We can implement a
similar technique to the one presented above to characterize an
unknown quantum optical process with only {\em one} weakly-entangled
state, i.e., a two-mode squeezed state (TMSS).

{\em Process tomography with one TMSS.---}
 A
TMSS is defined by $|\chi(q)\rangle=c_q \exp({q a^\dagger
b^\dagger})|00\rangle$, where $c_q=\sqrt{1-q^2}$, and $q$ is real.
The (un-normalized) maximally-entangled state here is
$|\Phi^+\rangle =\lim_{q\rightarrow 1} \exp(qa^\dagger
b^\dagger)|00\rangle=\sum_{k=0}^\infty |kk\rangle$, where
$x^\dagger$ is the creation operator for mode $x$.

We define the projection operator $\hat T(q) =c_q \exp[(\ln q)
a^\dagger a] $ which has the property: $ \hat T(q)\;(a,
a^\dagger)\;\hat T^{-1}(q)=(a/q,qa^\dagger)$. The TMSS
$|\chi(q)\rangle$ can be written as
\begin{equation}\label{tqlink}
\ket{\chi(q)}=\hat T(q)\otimes I \ket{\Phi^+}.
\end{equation}
Assume now that the black box process acts only on mode $b$ of the
bipartite state $|\chi(q)\rangle$. After the process, we obtain a
two-mode state $\Omega_q$.  We now wish to predict the evolution of
any state under the same process, using the information on how the
input state $|\chi(q)\rangle$ changes under this map. According to
Eq.~(\ref{tqlink}), we have
\begin{equation}
\Omega_q = \hat T(q) \otimes I\cdot \rho_{\varepsilon}\cdot \hat
T(q)\otimes I
\end{equation}
where $\rho_{\varepsilon}=I\otimes \varepsilon
\left(|\Phi^+\rangle\langle\Phi^+|\right)$. Naturally,
\begin{equation}\label{phiq}
\rho_{\varepsilon} = \hat T^{-1}(q) \otimes I \cdot  \Omega_q \cdot
\hat T^{-1}(q)\otimes I.
\end{equation}
We now also formulate the output state of any single-mode input
state $|\psi(\{c_k\})\rangle =\sum_k c_k |k\rangle$ of mode $b$.
According to Eq.~(\ref{e0}), we obtain the output state
\begin{eqnarray}\label{e12}
\rho_\psi = {\rm tr_a} \left[ |\psi^*(\{c_k/q^k\})\rangle\langle
\psi^*(\{c_k/q^k\})|\otimes I \cdot \Omega_q\right] \nonumber\\=
{_a\langle \psi^*(\{c_k/q^k\})| \Omega_q
|\psi^*(\{c_k/q^k\})\rangle_a}.
\end{eqnarray}
More explicit expressions can be obtained by using the $Q$-function.
If the single-mode input state on mode $b$ is a coherent state
$|\alpha\rangle$,  the output state  then becomes
$$\rho_{\alpha} =\langle \alpha^*|  \rho_{\varepsilon} |\alpha^*\rangle=
\langle \alpha^*|\hat T^{-1}(q)\otimes I \cdot \Omega_q \cdot \hat
T^{-1}(q)\otimes I|\alpha^*\rangle .
$$
Note that the state $|\alpha^*\rangle$ here
 is a single-mode coherent state on mode $a$.
Using the property of $\hat T(q)$ and the definition of  CSs,
$a|\alpha^*\rangle = \alpha^*|\alpha^*\rangle$, we easily find
\begin{equation}
\hat T^{-1}(q)\otimes I\;|\alpha^*\rangle  = \mathcal N_q(\alpha)\;
|\alpha^*/q\rangle
\end{equation}
where the factor $\mathcal N_q(\alpha)=
\exp\left[-|\alpha|^2(1-1/q^2)/2 \right] /c_q$, and
$|\alpha^*/q\rangle$ is a coherent state on mode $a$ defined by $a
|\alpha^*/q \rangle=(\alpha^*/q) |\alpha^*/q\rangle $. Thus, the
output state of mode $b$ is
\begin{equation}\label{e15}
\rho_{\alpha} = |\mathcal N_q(\alpha)|^2\;{_a\langle \alpha^*/q\;
|\Omega_q \;|\alpha^*/q\rangle_a}\;.
\end{equation}
Assume the $Q$-function for $\Omega_q$ is
$Q_{\Omega_q}(Z_a^*,Z_b^*,Z_a,Z_b)$. According to its definition,
$Q_{\Omega_q}(Z_a^*,Z_b^*,Z_a,Z_b)=\langle Z_a,Z_b|\Omega_q
|Z_a,Z_b\rangle$, where $|Z_a,Z_b\rangle$ is a two-mode coherent
state defined by
$(a,\;b)|Z_a,Z_b\rangle=(Z_a,\;Z_b)|Z_a,Z_b\rangle$. Hence the
corresponding density operator is
 $\Omega_{q} =
:Q(a^\dagger,b^\dagger,a,b): $\;, where the normal order notation
$:\ldots:$ indicates that any term inside it is reordered by placing
the creation operator in the left. For example, $:aba^\dagger
b^\dagger a:=a^\dagger b^\dagger a^2b$. Therefore, using
Eq.~(\ref{e15}) and the normally-ordered form of $\Omega_q$, we have
the following simple form for the $Q$-function
\begin{equation}\label{e16}
 Q_{\rho_\alpha}({Z_b}^*,Z_b) =
|N_q(\alpha)|^2Q_{\Omega_q}({\alpha}/{q},\;{Z_b}^*,\;{\alpha^*}/{q},\;
Z_b )
\end{equation}
of  the output state $\rho_{\alpha}$. Eqs.~(\ref{e15},\;\ref{e16})
are the explicit expressions of the output state for the input  of
{\em any} coherent state $|\alpha\rangle$. According to
Ref.~\cite{calgary}, if we know the output states for all input CSs,
we know the output states of all states in Fock space. In our
approach, given any input state $|\psi\rangle$, we can write it in
its linear superposition form in the coherent state basis, and then
obtain the $Q$-function of its output state by using
Eq.~(\ref{e16}).
These and Eqs.~(\ref{ddim},\;\ref{e12}), can be summarized as
follows:
\\{\bf Theorem 1}: Any process in Fock space is fully
characterized by the bipartite state $\Omega_q$, which is the output
of the initial TMSS $|\chi(q)\rangle$, if $q\not= 0$.  Any process
on $s$-dimensional states is characterized by the bipartite state
$\Omega_{\{r_k\}}$, which is the output state from the initial
bipartite state $|\phi(\{r_k\})$, if $r_k\not=0$ for all $k$s.

{\em Characterizing a Gaussian process by testing the map with a few
CSs.---} One can also choose to test a process with only single-mode
states. As shown in Ref.~\cite{calgary}, if we only use CSs in the
test, the tomography of  an unknown process in Fock space requires
tests with {\em all} CSs. Though this problem can be solved by
taking the photon-number-cut-off approximation, in a quantum-optical
process associated with intense light, one still needs a huge number
of different CSs for the test. Here we show that the most important
process in quantum optics, the Gaussian process, can be {\em
exactly}  characterized with only a few CSs in the test.

A Gaussian process maps Gaussian states into Gaussian states.
Therefore the $Q$-function of the operator $\rho_\varepsilon$ must
be Gaussian:
\begin{equation}\label{r0}
Q_{\rho_\varepsilon}(Z_{a}^*,Z_{b}^*,Z_a,Z_{b}) = \exp(c_0+L +
L^\dagger + S + S^\dagger + S_0 ),
\end{equation}
where $L=\mathcal G\left(
\begin{array}{c}Z_a\\Z_b\end{array}\right)$,
$S=\frac{1}{2}(Z_a,Z_b) X
\left(\begin{array}{c}Z_a\\Z_b\end{array}\right)$, $S_0=
(Z_a^*,Z_b^*)Y\left(\begin{array}{c}Z_a\\Z_b\end{array}\right)$,
$\mathcal G =(\Gamma_a,\Gamma_b)$, $ X = X^T=\left(\begin{array}{cc}X_{aa} & X_{ab}\\
X_{ba} & X_{bb}\end{array}\right)$, and $ Y =Y^\dagger =
\left(\begin{array}{cc}Y_{aa} & Y_{ab}\\ Y_{ba} & Y_{bb}
\end{array}\right) $. \\Before testing the map,
all these  are unknowns. The normally-ordered form of the density
operator $\rho_{\varepsilon}$ is $
:Q_{\rho_{\varepsilon}}(a^\dagger,b^\dagger,a,b):$\;.
The output state from any single-mode input coherent state $|
u\rangle $ (on mode $b$) is
\begin{equation}
\rho_{ u} = {\rm tr}_a\left[ \left(| u^*\rangle\langle u^*|\right)_a
\otimes I \cdot \rho_{\varepsilon}\right]
\end{equation}
Its $Q$-function is
\begin{eqnarray}\label{cohq}
Q_{\rho_{ u}}(Z_b^*,Z_b) = Q_{\rho_\varepsilon}( u, Z_b^*,
u^*,Z_b)\nonumber\\= \exp(c_u+L_{ u} + L_{ u}^\dagger + R +
R^\dagger + {R_0} ),
\end{eqnarray} where
$L_u = (\Gamma_b + u^*X_{ab} + uY_{ab}) Z_b$, $R = Z_b X_{bb}
Z_b/2$, ${R_0} = Z_b^* Y_{bb}Z_b$, and $c_u$ is determined by $c_0$,
$\Gamma_a$, $X_{aa}$ and $Y_{aa}$. Explicitly,
\begin{equation}\label{fator}
c_u =  c_0+Re\left(2\Gamma_a u^* +  u^* X_{aa} u^* +
uY_{aa}u^*\right)
\end{equation}
The quadratic functional terms, ($R,\;R^\dagger,\;R_0$) on the
exponent in Eq.~(\ref{cohq}) are independent of $u$; these terms
must be the same for the output states from any initial CSs.
Therefore, these can be known by testing the map with  one coherent
state. Thus, we do not need to consider these terms below.
 Now suppose that we test the process with six different CSs,
$|\alpha_i\rangle$, and $i=1,\cdots 6$. Assume also that the
detected $Q$-function of the output states is
\begin{equation}\label{qa}
Q_{\rho_{\alpha_i}}(Z^*_b,Z_b)  = \exp(c_i+D_{i} + D_{ i}^\dagger +
R + R^\dagger + {R_0} )
\end{equation}
where $D_i = d_i Z_b$ is the detected (hence known) linear term.
According to Eq.~(\ref{cohq}), the $Q$-function of the output state
from the initial state $|\alpha_i\rangle$ of mode $b$ must be
$Q_{\rho_{\alpha_i}}(Z^*_b,Z_b) = Q_{\rho_\varepsilon}( \alpha_i,
Z_b^*, {\alpha_i}^*,Z_b)$. Therefore,  we can derive self-consistent
equations  by using the detected data from $\rho_{\alpha_i}$ and
setting $u=\alpha_i$ in Eq.~(\ref{cohq}):
 \begin{equation}\label{self}
  L_i=D_i;\: c_{\alpha_i} = c_i
\end{equation}
 where $L_i $, $c_{\alpha_i}$ are just $L_u$, $c_u$, respectively, after setting $u=\alpha_i$
 in Eqs.~(\ref{cohq}-\ref{fator}); $D_i$ and $c_i$ are known from tests.
 Explicitly, $L_i=(\Gamma_b + {\alpha_i}^*X_{ab}
+ {\alpha_i}Y_{ab}) Z_b$. The first part of Eq.~(\ref{self}) causes:
\begin{eqnarray}\label{KK}
K\cdot
 \left( \Gamma_b,\; X_{ab},\;Y_{ab} \right)^T = d,
\end{eqnarray}
where $K=\left(
\begin{array}{ccc}
1 & {\alpha_1}^* & \alpha_1\\
1 & {\alpha_2}^* & \alpha_2\\
1 & {\alpha_3}^* & \alpha_3
\end{array}
\right) $, $d=\left(d_1,\;d_2,\;d_3\right)^T$. There are three
unknowns  ($\Gamma_b$, $X_{ab}$, and $Y_{ab}$) with three equations
now. We find
\begin{equation}\label{root}
\left( \Gamma_b,\; X_{ab},\;Y_{ab} \right)^T = K^{-1}d\;.
\end{equation}
If the Gaussian process is known to be trace-preserving, then
Eq.~(\ref{root})  completes the tomography: up to a numerical
factor, we can deduce all the output states of the other input CSs,
$|\alpha_i\rangle$, for $i=4,5,6$. The term $c_i$ can be fixed
through normalization, which is determined  by the quadratic and
linear functional terms on the exponent of the $Q$-functions.
Knowing these $\{c_i\}$, one can construct $\rho_{\varepsilon}$
completely, as shown below.  For any map, $c_i$ can be known from
tests with $|\alpha_i\rangle$.  We then have
\begin{equation}\label{JJ}
J\cdot \left( c_0,\;  \Gamma_a,\;  \Gamma^*_a,\;  X_{aa},\;
X^*_{aa},\;  Y_{aa} \right)^T=c\;,
\end{equation}
$J=\left(\begin{array}{cccccc} 1 & \alpha^*_1 & \alpha_1 & {1\over
2}\alpha^{*2}_1 & {1\over 2}\alpha^2_1 & |\alpha_1|^2 \\ 1 &
\alpha^*_2 & \alpha_2 & {1\over 2}\alpha^{*2}_2 & {1\over
2}\alpha^2_2 & |\alpha_2|^2 \\ 1 & \alpha^*_3 & \alpha_3 & {1\over
2}\alpha^{*2}_3 & {1\over 2}\alpha^2_3 & |\alpha_3|^2 \\ 1 &
\alpha^*_4 & \alpha_4 & {1\over 2}\alpha^{*2}_4 & {1\over
2}\alpha^2_4 & |\alpha_4|^2 \\ 1 & \alpha^*_5 & \alpha_5 & {1\over
2}\alpha^{*2}_5 & {1\over 2}\alpha^2_5 & |\alpha_5|^2 \\ 1 &
\alpha^*_6 & \alpha_6 & {1\over 2}\alpha^{*2}_6 & {1\over
2}\alpha^2_6 & |\alpha_6|^2 \end{array}\right)$,
$c=\left(\begin{array}{c} c_1 \\ c_2\\ c_3\\c_4\\c_5\\c_6
\end{array}\right)$ \\for the second part of Eq.~(\ref{self}). Thus
\begin{equation}\label{rootJ}
\left( c_0,\;  \Gamma_a,\;  \Gamma^*_a,\;  X_{aa},\; X^*_{aa},\;
Y_{aa} \right)^T=J^{-1}c\; .
\end{equation}
 {\bf Theorem 2}: Given $K$ and $J$ defined by
Eqs.~(\ref{KK},\;\ref{JJ}), then the QPT of any single-mode Gaussian
process in Fock space can be performed   with six input CSs, when
$\det K\not=0$ and $\det J\not=0$. The QPT of any trace-preserving
single-mode Gaussian process in Fock space can be executed  with
three input CSs, when $\det K\not=0$.

For example, one can simply choose $\alpha_1=0$, $\alpha_2=1$,
$\alpha_3=i$, $\alpha_4=-1$, $\alpha_5=-i$, and $\alpha_6=1+i$. One
finds
\begin{eqnarray}\label{28}
c_0&=&c_1,\;\Gamma_b=d_1 \nonumber \\
X_{ab}&=&\left[-{(1+i) }d_1+d_2+{i }d_3\right]/2 \nonumber \\
Y_{ab}&=&\left[-{(1-i) }d_1+d_2-{i}d_3\right]/2 \nonumber \\
\Gamma_a&=&\left(c_2+ic_3-c_4-{i }c_5\right)/4
\nonumber \\
X_{aa}&=&\left[{2i }c_1+{(1-2i) }c_2-{(1+2i) }c_3+c_4-c_5+{2i }c_6\right]/4 \nonumber\\
Y_{aa}&=&-c_1+(c_2+c_3+c_4+c_5)/4
\end{eqnarray}
where $\{d_i\}$  and $\{c_i\}$ are defined in Eq~(\ref{qa}).

{\em An example.---}As  a check of  our conclusion, we calculate the
output state of a beam-splitter (BS) process as shown in Fig.~1. The
BS has input modes $b$ and $c$ and output modes  $b'$ and $c'$.
Regarding this as a black-box process, the only input is mode $b$
and the only output is mode $b'$.  We set mode $c$ to be vacuum. The
BS transforms  the creation operators  of modes $b$, $c$ by:
\begin{equation}\label{change}
U_{BS}\left(b^\dagger,\;  c^\dagger
\right)U^{-1}_{BS}=\left(b^\dagger,\;  c^\dagger \right)\;M_{BS}
\end{equation}
where $M_{BS}=\left(\begin{array}{cc}\cos\theta &\sin\theta
 \\ -\sin\theta &\cos\theta
\end{array}\right)$.
If we test such a process with a coherent state $|\alpha_i\rangle$,
we shall find $\rho_{\alpha_i}= |\cos\theta
\alpha_i\rangle\langle\cos\theta\alpha_i|$. Comparing this with
Eq.~(\ref{qa}), we have $d_i=\alpha^*_i\cos\theta$ and
$c_i=-|\alpha_i\cos\theta|^2$. Using
Eqs.~(\ref{root},\;\ref{rootJ}), we find
\begin{eqnarray}
Y_{bb}=-1 ,\; X_{ab}=\cos\theta,\;
Y_{aa}=-\cos^2\theta,\;\nonumber\\
\Gamma_a=\Gamma_b=Y_{ab}=X_{aa}=X_{bb}=c_0=0
\end{eqnarray}
Therefore $\rho_{\varepsilon} =:\exp(a^\dagger b^\dagger \cos\theta
-a^\dagger a \cos^2\theta-b^\dagger b+a b \cos\theta):$ . With this
we can predict the output state of {\em any} input state, for
example the displaced squeezed state  $
\ket{\xi(r,Z)}=\exp\left({-{r \over 2}b^{\dagger 2}+{r \over
2}b^2}\right)\exp\left({Z b^\dagger-Z^* b }\right)\ket{0} $, where
$r$ is real. According to our Eq.~(\ref{e0}),  $ \rho_\xi ={\rm
tr}_a\left[(\ket{\xi(r,Z^*)}\bra{\xi(r,Z^*)})_a \otimes I_b \cdot
\rho_{\varepsilon}\right]. $ As a result,
$Q_{\rho_\xi}(Z^*_b,Z_b)=C\exp(\mathcal H_1 -\mathcal H_2 + \mathcal
H_3 + \mathcal H_4)$, where $C$ is the normalization factor, and
$\mathcal H_1 =|Z_b|^2(\tanh^2 r\sin^2 \theta-1 )/g ,\; \mathcal
H_2= (Z^2_b+{Z_b^*}^2)\tanh r\cos^2\theta /(2g),\; \mathcal H_3
=Z_b\cos\theta (Z^*-Z\tanh r\sin^2\theta)/(g \cosh r),\;\mathcal
H_4=Z^*_b \cos\theta (Z-Z^*\tanh r\sin^2\theta)/(g \cosh r)$, and
$g=1-\tanh^2 r\sin^4\theta $. This is same with the result from
direct calculations using Eq.~(\ref{change}).
\begin{figure}\label{f1}
  \includegraphics[width=150pt]{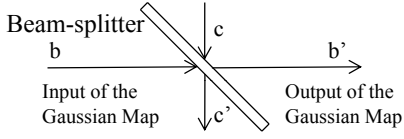}\\
  \caption{Gaussian Map constructed by a beam-splitter}\label{BS}
\end{figure}

{\em Multi-mode extension.---}  Multi-mode Gaussian QPT has many
important applications. For example, it applies to a complex linear
optical circuit  with BSs, squeezers, homodyne detections, linear
losses, Gaussian noises and so on. Consider now a Gaussian process
acting on a $k$-mode input state (on mode $b_1,b_2,\cdots b_k$),
with outcome also a $k$-mode state. Even though other
methods~\cite{calgary} can also be extended to the multi-mode case,
the number of input states required there increases exponentially
with the number of modes $k$, because the number of ket-bra
operators
 $|\{n_i\}\rangle\langle \{m_i\}|$ in Fock space increases
exponentially with $k$. As shown below, the number of input states
in our method increases {\em polynomially}.

To apply  isomorphism~\cite{liusky}, we consider $k$ pairs of
maximally entangled states, each on modes $a_1,b_1$;
$a_2,b_2$,$\cdots$ $a_k,b_k$. Explicitly, $ | \Phi^+\rangle =
|\phi^+\rangle_1|\phi^+\rangle_2\cdots|\phi^+\rangle_k. $ Here
$|\phi^+\rangle_i=\lim_{q\rightarrow 1} \exp\left({qa_i^\dagger
b_i^\dagger}\right)|00\rangle$ indicates a maximally-entangled state
on modes $a_i,\;b_i$.  Subspaces $a$ and $b$ each are now $k$-mode.
Any state $|\psi\rangle$ in subspace $b$, can still be written in
the form of Eq.~(\ref{00}), with the new definitions for
$|\psi\rangle$ and $|\Phi^+\rangle$. Using Eq.~(\ref{phiq}), it is
obvious that the output state of these $k$-pairs-TMSS fully
characterize the process. A $k$-mode QPT can also be tested with
$k$-mode CSs, if the process in Gaussian. The main
Eqs.~(\ref{root},~\ref{rootJ}) still hold after redefining the
notations there. First, $\Gamma_a,\;\Gamma_b$, $u$, $\alpha_i$,
$d_i$, $Z_a,Z_b$ are now  $k$-mode vectors. For example,
$|\alpha_i\rangle=|\alpha_{i1},\alpha_{i2},\cdots\alpha_{ik}\rangle$,
$Z_b = (Z_{b1},Z_{b2},\cdots Z_{bk})$, $d_i = (d_{i1}, d_{i2},\cdots
d_{ik})$, and so on. Following Eq.~(\ref{r0}), $\mathcal {X}_{xy}$
is now a $k\!\times\! k$ matrix, for $\mathcal X = X$ or $Y$ with
$x=a,b$ $y=a,b$.  We still apply Eqs.~(\ref{root},\;\ref{rootJ}) to
calculate \{$\Gamma_b$, $X_{ab},\; Y_{ab}$\} and \{$\Gamma_a$,
$X_{aa}$, $Y_{aa}$\}, respectively, but keep in  mind that the
matrices $K$, $J$ and symbols $d$, $c$ are now redefined. There are
$(2k +1)k$ unknowns in ($\Gamma_B,\; X_{ab},\; Y_{ab}$). We need
$2k+1$ different CSs of $k$-mode to fix these unknowns. Matrix $K$
is now $(2k+1)\times (2k+1)$, since each $\alpha_i$ here is a
$k$-mode row vector. Here $d$ is a $(2k+1)\times k$ matrix as $d^T=
\left(d_1^T,\;d_2^T, \cdots d_{2k+1}^T \right)$, with $d_i =
(d_{i1},d_{i2}, \cdots d_{ik})$. Similarly, $J$ is now $N\times N$
and $N=(k+1)(2k+1)$, since $\alpha_i^2$ and $|\alpha_i|^2$ here are
row vectors of $\alpha^2_i=(E_{i1},\;E_{i2},\cdots E_{ik}) $ and
$|\alpha_i|^2=(\tilde E_{i1},\;\tilde E_{i2},\cdots \tilde E_{ik})$,
 and each element of $E_{im}$ (or $\tilde E_{im}$) is a vector
 with ($k-m+1$) modes (or $k$ modes), as
 $E_{im} = (\alpha_{im}^2,\alpha_{im}\alpha_{i,m+1},\alpha_{im}\alpha_{i,m+2},\cdots \alpha_{im}\alpha_k,\alpha_k^2)$
 and
 $\tilde E_{im}= (\alpha_{im}\alpha^*_{i1},\alpha_{im}\alpha^*_{i2},\cdots\alpha_{im}\alpha^*_{i,k-1},\alpha_{im}\alpha^*_{ik})$.
 Obviously, $c$ is a column vector with $N$ elements.
 Therefore we conclude with this:
\\{\bf Corollary 1}: Any $k$-mode map $\varepsilon$ in Fock space is characterized
by the output state of $k$-pair-TMSS under one-sided map $I\otimes
\varepsilon$.  Any $k$-mode Gaussian QPT can be performed with
$(k+1)(2k+1)$ different CSs of $k$-mode; or with $2k+1$ different
CSs of $k$-mode if the process is trace-preserving.

In summary, we have presented explicit formulas quantum process
characterization with only one weakly entangled state, as well as
the tomography of a quantum optical Gaussian process  with  a few
different coherent states. These results have been extended to
multi-mode quantum optical process and the number of test states
required increases only polynomially with the number of modes.

{\bf Acknowledgments}
 XBW is supported by the NSFC under Grant No.~60725416, the National
Fundamental Research Programs of China Grant No. 2007CB807900 and
2007CB807901, and China Hi-Tech Program  Grant No.~2006AA01Z420. FN
acknowledges partial support from the NSA, LPS, ARO, DARPA,  NSF
Grant No.~0726909, JSPS-RFBR Contract No.~09-02-92114, Grant-in-Aid
for Scientific Research (S), MEXT Kakenhi on Quantum Cybernetics,
and FIRST (Funding Program for Innovative R\&D on S\&T).


\vspace*{-0.1in}
\begin{thebibliography}{99}
\vspace*{-0.1in} \vspace*{-0.1in}
\bibitem{qpt}J.F. Poyatos, J.I. Cirac, and P. Zoller, Phys. Rev.
Lett. {\bf 78}, 390 (1997); G.M. D'Ariano and P. Lo Presti, Phys
Rev. Lett. {\bf 86}, 4195 (2001); M. Mohseni, A.T. Rezakhani, and
D.A. Lidar, Phys. Rev. A {\bf 77}, 032322 (2008).
\bibitem{calgary}K. Lobino et al, Science {\bf 322}, 563 (2008); S. Rahimi-Keshari et al, arXiv:1009.3307v1.
\bibitem{liusky}A. Jamiolkowski, Rep. Math. Phys, {\bf 3}, 275 (1972).
%
\bibitem{gie}J. Eisert and M.B. Plenio, Phys. Rev. Lett. {\bf 89}, 097901
(2002); G. Giedke and J.I. Cirac, Phys. Rev. A {\bf 66}, 032316
(2002).
\end{thebibliography}
\end{document}